 \documentclass[preprint,showpacs,preprintnumbers,amsmath,amssymb]{revtex4}

\usepackage{graphicx}
\usepackage{dcolumn}
\usepackage{bm}

\begin{document}
 

\title{Topology studies of hydrodynamics using two particle correlation analysis}

\author{J.~Takahashi}\email{jun@ifi.unicamp.br}\affiliation{Universidade Estadual de Campinas, S\~{a}o Paulo, Brazil}
\author{B.M.~Tavares}\affiliation{Universidade Estadual de Campinas, S\~{a}o Paulo, Brazil}
\author{W.~L.~Qian}\affiliation{Universidade de S\~{a}o Paulo, S\~{a}o Paulo, Brazil}
\author{R.~Andrade}\affiliation{Universidade de S\~{a}o Paulo, S\~{a}o Paulo, Brazil}
\author{F.~Grassi}\affiliation{Universidade de S\~{a}o Paulo, S\~{a}o Paulo, Brazil}
\author{Y.~Hama}\affiliation{Universidade de S\~{a}o Paulo, S\~{a}o Paulo, Brazil}
\author{T.~Kodama}\affiliation{Universidade Federal do Rio de Janeiro, Rio de Janeiro, Brazil}
\author{N.~Xu}\affiliation{Lawrence Berkeley National Laboratory, Berkeley, California, USA}


\begin{abstract}
The effects of fluctuating initial conditions are studied in the context of relativistic heavy ion collisions
where a rapidly evolving system is formed.
Two particle correlation analysis is applied to events generated with the NEXSPHERIO
hydrodynamic code, starting with fluctuating non-smooth initial conditions (IC).
Results show that the non-smoothness in the IC survives the hydro-evolution and can
be seen as topological features of the angular correlation function of the particles emerging 
from the evolving system.
A long range correlation is observed in the longitudinal direction and
in the azimuthal direction a double peak structure is observed in the opposite direction to the trigger particle.
This analysis provides clear evidence that these are signatures of the combined effect of tubular structures 
present in the IC and the proceeding collective dynamics of the hot and dense medium. 

\end{abstract}


\maketitle

Relativistic heavy ion collisions create a rich environment to study the behavior of matter under extreme conditions. 
QCD degrees of freedom and collective behavior have to be considered to explain the different features observed in the experimental data. 
During the dynamic evolution of the fireball created in these collisions, different mechanisms compete and the features observed in the final particles are a complex result of this competition, thus, the study of the experimental data requires the development of analysis tools and detailed comparison to theory and phenomenological models. 
Within this context, two particle correlation analysis have allowed to observe the suppression of the Jets by the medium~\cite{ref:star1} and also the appearance of a long range angular correlation in the longitudinal direction~\cite{ref:star2,ref:phenix}, also known as ``The Ridge''.
Different models have been suggested for the ``Ridge''. Our work has some points in common with the general argument presented in~\cite{ref:voloshin,ref:Shuryak} and further developped in the framework of Glasma flux tubes~\cite{ref:LarryRaju,ref:LarryGavin}.
However, to isolate signals from specific physical mechanisms it is necessary to do a careful subtraction of other global features such as the effects generated by collective behaviors.
In relativistic heavy ion collisions, collective behavior of the system has been successfully described with the application of hydrodynamic descriptions.
We have studied two particle correlation analysis using data generated with a complete hydrodynamic model to obtain some insight on the effects of hydro-evolution and possibly shed some light into some of the observed features in the experimental data.

\begin{figure}[hbt]
\includegraphics[width=8.5cm]{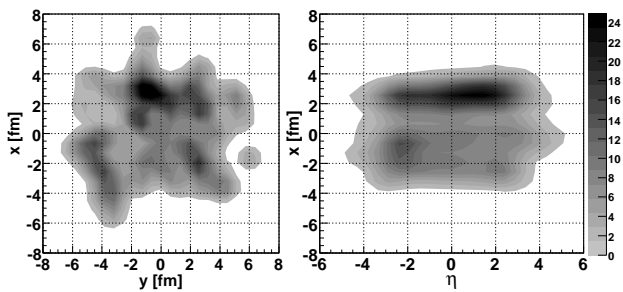}
\caption{\label{fig1} Example of a transverse and longitudinal profile of the initial energy density distribution in $GeV/fm^{3}$ generated using the NEXUS code~\cite{ref:nexus}. This is equivalent to an event of $Au+Au$ collisions at $\sqrt{s_{NN}} = 200 GeV$ with centrality of top $10\%$.}
\end{figure}

In this work, the reaction dynamics is calculated through the NEXSPHERIO code -- a combination of the event generator NEXUS~\cite{ref:nexus} and the hydrodynamical code SPHERIO~\cite{ref:spherio}. 
NEXUS is a code based on the Gribov-Regge model of hadronic collisions and provides the initial conditions (IC) for the hydrodynamical evolution.
For each event, the IC are defined as a set of spatial three dimensional distributions such as energy density, baryon number density and velocity fields. 
In figure~\ref{fig1} we show an example of a single event energy density profile in the transverse (left) and longitudinal plane (right).
This particular event shown in figure~\ref{fig1} is equivalent to a central $Au+Au$ collisions at $\sqrt{s_{NN}} = 200 GeV$.
The darker regions observed in the density profiles are generated by a higher density of flux-tubes in the IC.
The system then evolves through a sequence of locally equilibrated states, where the dynamics of these states are determined by an equation of state. 
In the SPHERIO code, hydrodynamical equations of motion -- that represent energy-momentum conservation laws -- are solved in 3 dimensions using the variational relativistic Smoothed Particle Hydrodynamics (SPH) method \cite{ref:sphmethod}, without the use of any symmetry simplifications. 
More details can be found in \cite{ref:spherio2, ref:wliangeos}. 
After some time of the evolution the local thermal equilibrium pre-condition is no longer valid and decoupling criterion is employed when hadrons are then generated through the Cooper-Frye procedure~\cite{ref:cooperfrye}.
In the final part of our simulation code particles that have short life times are decayed, thus the final result of our code are particles equivalent to the ones that can be measured in the experiments.
It is important to note that, within this model the energy momentum tensor used as the IC contains the contributions from both soft and hard particles and is interpreted as thermalized. Thus, particles with high transverse momentum ($p_{T}$) observed in the final part of the simulation originate from the tails of the boosted thermal distributions at the end of the hydro simulation.

We have generated on the order of 200,000 events of $Au+Au$ collisions at $\sqrt{s_{NN}}=200 GeV$ with two different centrality classes. 
The first set of events was generated with impact parameters of central collisions that corresponds to the upper $10\%$ of the total cross section and the second set was generated considering impact parameters of peripheral collisions with a cross section fraction from $40\%$ to $60\%$. Only the charged particles were considered in this analysis to allow for a comparison to the experimental data available from the RHIC experiments.

Two particle correlation analysis was applied to the data generated by the NEXSPHERIO code using similar methods as used by the STAR experiment~\cite{ref:star2}. 
In this method, particles with $p_{T}$ higher than a threshold are classified as trigger particles and the angular difference of the other particles in the same event, called here as the associated particles, are calculated with respect to the direction of each trigger particle in the azimuthal direction $\Delta \phi$ and in the longitudinal direction $\Delta \eta$.
This method was originally applied to find angular correlations due to particle Jets in heavy ion collisions. 
For this reason, to reduce the contribution of the thermally produced particles and enhance the signal from the Jets, a low $p_{T}$ threshold was also applied for the associated particles.
Typical values of the low $p_{T}$ threshold was around $2 GeV/c$. 
Since in this analysis the main interest is in the topology of the structures created by the hydrodynamic evolution, the associated particle low $p_{T}$ threshold was reduced to $1 GeV/c$.
The trigger particle $\eta$ acceptance was limited to $\pm 1.5$ to increase the similarity with the analysis method used by the experiments. 
Analysis without the $\eta$ limitation of the trigger particle was also tested and results show no qualitative difference with the results shown here.

\begin{figure}[hbt]
\includegraphics[width=8.4cm]{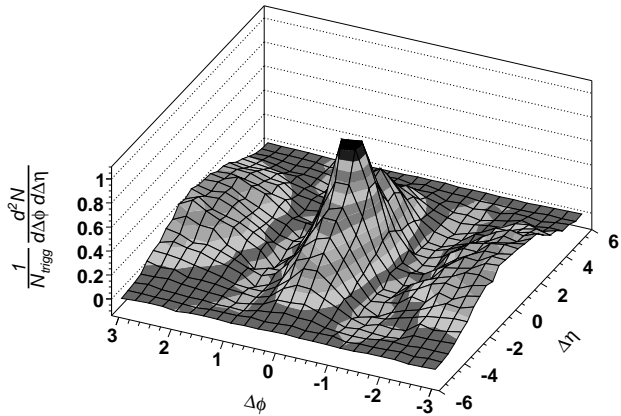}
\caption{\label{fig2} Two particle correlation in $\Delta \eta$ and $\Delta \phi$, for simulated data with equivalent centrality of $0-10\%$, after subtracting the event mixing. In this case we used the trigger particle requirement of $p_{T}>2.5 GeV/c$ and associated particle requirement of $p_{T}>1.0 GeV/c$. }
\end{figure}

In the longitudinal direction, due to the shape of the $\eta$ distribution of the particles, the $\Delta \eta$ correlation distribution shows a strong variation that is higher for small $\Delta \eta$ angles and decrease for higher values of $\Delta \eta$. 
This structure in the correlation function is subtracted using event mixing technique, where the same correlation histograms of $\Delta \eta$ and $\Delta \phi$ are generated using trigger particles from one event and associated particles from different events. 
The event mixing is done within the same event centrality class, thus with events that have similar particle multiplicities.
Figure~\ref{fig2} shows the result of the two particle correlation function in $\Delta \eta$ and $\Delta \phi$ after the mixed event subtraction, calculated for central events. 
In this case, the correlation function was calculated using trigger particle threshold of $p_{T}>2.5 GeV/c$ and associated particle requirement of $p_{T}>1.0 GeV/c$. 
There is a clear structure in the topology of the correlation distribution with a variation in the $\Delta \phi$ direction. 
The $\Delta \phi $ correlation has the contribution from the anisotropy parameter ($v_{2}$) that adds a cosine type oscillation in the $\Delta \phi$ direction. 

\begin{figure}[hbt]
\includegraphics[width=8.4cm]{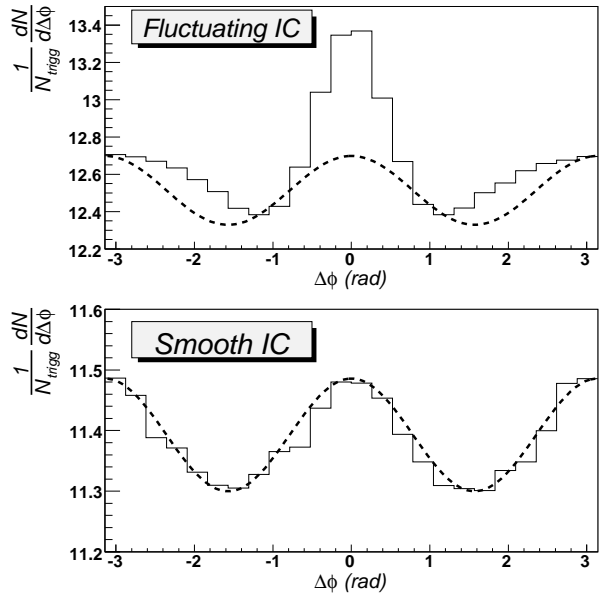}
\caption{\label{fig3} Projection in $\Delta \phi$, for the $\Delta \eta$ range between $\pm 0.5$. Curve in dashed line is the flow contribution normalized using ZYAM (Zero Yield At Minimum)~\cite{ref:zyam} method. This is from top $10\%$ $Au+Au$ collisions at $\sqrt{s_{NN}} = 200 GeV$.}
\end{figure}

Figure~\ref{fig3} shows the projection of the correlation distribution in the $\Delta \phi$ direction, integrating the $\Delta \eta$ range between 
$\pm 0.5$.
The top plot corresponds to the projection histogram of the central events and the dashed curve represent the $v_{2}$ contribution normalized using ZYAM (Zero Yield At Minimum)~\cite{ref:zyam} method. 
The $v_{2}$ dependency on the $p_{T}$ was calculated based on the method as described in~\cite{ref:spheriov2}.
The amplitude of the $v_{2}$ contribution is adjusted based on the mean $v_{2}$ value of the trigger particles and the associated particles and the pedestal offset is adjusted equalizing the minimum yield of the correlation function with the flow contribution.

It is clear from the comparison between the $\Delta \phi$ projection and the flow contribution that there is an excess of the correlation yield suggesting that the topology observed cannot be explained considering just the $v_{2}$ contribution. 
Same comparison was done also for events generated with the SPHERIO code using a smooth IC (without the fluctuations), and the equivalent result is shown in the bottom part of figure~\ref{fig3}. 
The smooth IC were generated averaging over several NEXUS events.
In this case, the $v_{2}$ curve agrees with the particle correlation function indicating that the anisotropy $v_{2}$ is the only contributor to the topology of the two particle correlation function. 

\begin{figure}[hbt]
\includegraphics[width=8.4cm]{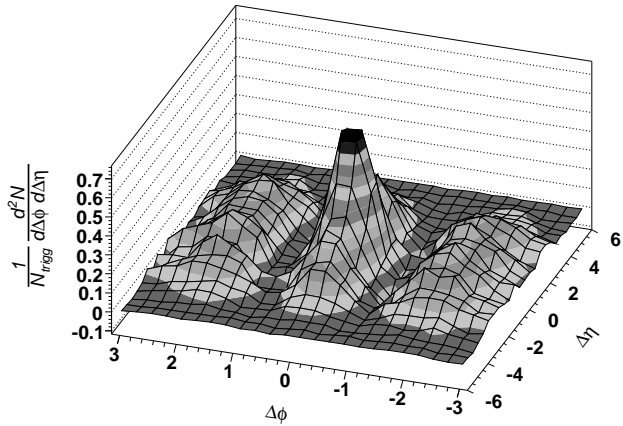}
\caption{\label{fig4}Two particle correlation in $\Delta \eta$ and $\Delta \phi$, for simulated central data, after subtracting the event mixing and flow contribution. Flow contribution was calculated using the ZYAM method and a mean v2 calculated for the entire $\eta$ range.}
\end{figure}

Figure~\ref{fig4} shows the topology of the correlation function after subtracting the $v_{2}$ contribution using the ZYAM method applied to the $\Delta \phi$ projection of each interval of $\Delta \eta$. 
It is clear that there is still an excess of the correlation yield in the near side of the trigger particle ($\Delta \phi \approx 0$) and also some excess in the away side ($\Delta \phi \approx \pi$). 
It can also be noted that in the near side, the correlation function is narrow in the $\Delta \phi$ but extends over several units of pseudo-rapidity. In the away side, the correlation function also extends over several units of rapidity with two ``Ridge'' like structures peaked at $\Delta \phi \approx \pm 2$.

\begin{figure}[hbt]
\includegraphics[width=8.4cm]{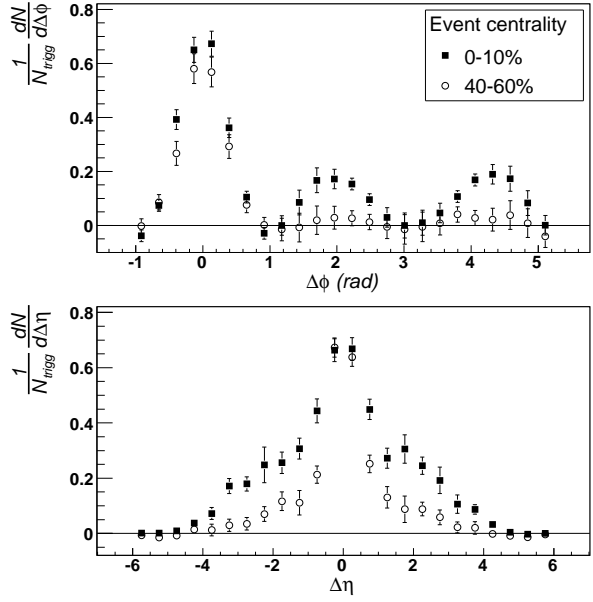}
\caption{\label{fig5} Top plot shows the $\Delta \phi$ projection for the simulated data equivalent to central collisions (solid squares) and peripheral collisions (open circles). Bottom plot shows the equivalent $\Delta \eta$ distribution, for the near side ($-1.0 > \Delta \phi >1.0$) for central (Solid squares) and peripheral (open circles) collisions. The error bars include the statistical error of each bin and the error due to v2 of the ZYAM method.}
\end{figure}

Figure~\ref{fig5} top plot shows the $\Delta \phi$ projection of the two particle correlation, integrated over $\Delta \eta$ range between $\pm 1$, for events from the central (solid squares) and peripheral (open circles) centrality classes. 
The horizontal axis was shifted by $\pi / 2$ to help in the visualization and distinguishing the near side and away side. 
Both data sets show a clear narrow correlation peak in the near side.
In the away side, the central data show a double peak structure with a dip at $\Delta \phi=\pi$ that appears after the subtraction of the $v_{2}$ contribution while the peripheral data show no clear correlation in the away side. 
The double peak structure in the away side was observed in the experimental data and is the focus of several different explanations such as away side Jet deflection and coherent conical emission due to Jet energy loss\cite{ref:machcone}.
In our model, we do not have a particle Jet traversing the medium, but rather the evolution of IC hot spots through the hydrodynamic expansion, which generate a topology structure that can be observed in the characteristics of the final particle correlation functions.
The shape of the away side structure in the $\Delta \phi$ is strongly dependent on the $v_{2}$ values. Further investigation of the origin of this structure and the dependency on the $v_{2}$ calculation is in progress.

The $\Delta \eta$ projection of the near side correlation is shown in the second plot of figure~\ref{fig5} for the two different centrality classes, central events shown with solid squares and peripheral events with open circles. 
The uncertainties in each point include the statistical error of each bin and the uncertainty in the $v_{2}$ value that was subtracted. 
It is clear that the residual correlation extends over several pseudo-rapidity units, but the range of the correlation in the longitudinal direction seems to decrease with the event centrality. 
This ``Ridge'' structure in the $\Delta \eta$ correlation was also observed in the experimental data at RHIC by the STAR ~\cite{ref:star2} and PHENIX experiments~\cite{ref:phenix}. 
The observed feature extends up to two units of $\Delta \eta$ within the limits of the detector acceptance and has a strong peak that corresponds to the near side Jet correlation on top of the ``Ridge'' structure.

Our analysis show that the large extension of the correlation in the longitudinal direction is related to the extension of the higher energy density regions in the IC. 
This suggests that the initial non-smoothness of the system survives the hydrodynamic evolution and can be observed in the final experimentally observed topology in the two particle correlation function.
It is also necessary to evaluate the effects of final state hadronic interactions to the shape of the topology structures observed here.
Recent results considering hydro-evolution calculations coupled with hadronic cascade model~\cite{ref:Hirano} shows that the effect to the hydrodynamic features such as $v_{2}$ is small, thus should not affect the main features observed here. 
Correlation functions from events generated using only the NEXUS code without the hydrodynamic evolution was also verified and we did not observe any type of topology structure in the correlation function, except for a narrow Jet like peak structure in $\Delta \eta =0$ and $\Delta \phi =0$.
Events generated considering just pure hydrodynamics starting with smooth IC also do not generate the topology structures.
Only when we couple the NEXUS outputs (the IC) with the SPHERIO calculation (the hot and dense medium) {\it in an event-by-event fashion}, the ``Ridge'' structure can be observed.

Thus, in conclusion, the topology structures observed in this analysis are due
to the combined effect of the fluctuating non-smooth initial conditions and the proceeding
collective dynamics of the hot and dense medium created in the heavy ion collisions.
It is worthy to note that what we observed in this analysis may share similarities in other fields. 
For exemple, in astrophysics~\cite{ref:Smoot}, the super horizon fluctuations in the Cosmic Microwave Background are thought to have originated from the small quantum fluctuations present at the time of inflation in the early universe. Also, if indeed the initial conditions in the heavy ion collisions manifest themselves through the two-particle correlations, this opens the possibility to study particle production mechanisms on the sub-fermi scale. In addition, the SPH method used in this work was originally developed in astrophysics~\cite{ref:Monaghan} with various applications.
In the context of particle interactions in heavy ion collisions, our results provide a clear picture for the dynamic 
evolution at the foremost early stage of the collisions.

We wish to thank Dr. Larry McLerran, Dr. Klaus Werner, Dr. Tetsufumi Hirano and Dr. Paul Sorensen for fruitful discussions. This work received support from Funda\c{c}\~{a}o de Amparo a Pesquisa do Estado de S\~{a}o Paulo, FAPESP, Comiss\~{a}o Nacional de Pesquisa e Desenvolvimento CNPq, Funda\c{c}\~{a}o de Amparo a Pesquisa do Estado do Rio de Janeiro, FAPERJ, PRONEX of Brazil and the U.S. Department of Energy under Contract No. DE-AC03-76SF00098.

\end{document}